\documentclass[preprint, 11pt, a4paper]{aastex63}
\usepackage[english]{babel}

\usepackage{amsmath} 

\usepackage{graphicx}
\usepackage{natbib}
\usepackage{gensymb}

\newcommand{\lya}{Lyman-$\alpha$~{}}

\begin{document}
\title{Radiation pressure acting on the neutral He atoms in the Heliosphere.}

\correspondingauthor{I. Kowalska-Leszczynska}
\email{ikowalska@cbk.waw.pl}

\author[0000-0002-6569-3800]{I. Kowalska-Leszczynska}
\affil{Space Research Centre PAS (CBK PAN),\\
Bartycka 18A, 00-716 Warsaw, Poland} 

\author[0000-0002-5204-9645]{M. A. Kubiak}
\affil{Space Research Centre PAS (CBK PAN),\\
Bartycka 18A, 00-716 Warsaw, Poland}

\author[0000-0003-3957-2359]{M. Bzowski}
\affil{Space Research Centre PAS (CBK PAN),\\
Bartycka 18A, 00-716 Warsaw, Poland}

\begin{abstract}
The Interstellar Neutral Helium (ISN He) is an important source of information on the physical state of the Local Interstellar Medium.
Radiation pressure acting on the neutral helium atoms in the heliosphere has always been neglected, its effect has been considered insignificant compared to gravitational force.
The most advanced numerical models of ISN He take into account more and more subtle effects, therefore it is important to check if the effect of radiation pressure is still negligible.
In this paper, we use the most up-to-date version of the Warsaw Test Particle Model (WTPM) to calculate the expected helium distribution in the heliosphere, and simulate the flux of ISN He observed by the Interstellar Boundary Explorer (IBEX) and in the future by the Interstellar Mapping and Acceleration Probe (IMAP).
We compare results calculated with and without radiation pressure during low and high solar activity.
The results show that in the analysis of IBEX-Lo observations the radiation pressure acting on typical helium causes flux differences at a level of 1-4\% and is comparable to the observational errors. 
For the more sensitive IMAP-Lo instrument, there are some regions in the considered observations configurations where radiation pressure causes potentially statistically significant changes in the calculated fluxes.
The effect can be up to 9\% for the indirect beam and is likely to be higher than the estimated errors.
Therefore, we claim that in the future analysis of the IMAP-Lo observations radiation pressure acting on ISN He should be considered.
\end{abstract}

\section{Introduction}
\label{sec:intro}
\noindent

Helium, following hydrogen, is the most important element that affects the shape and properties of the heliosphere.
Neutral atoms entering the heliosphere through the heliopause move along trajectories determined by the gravitational force of the Sun.
During their journey, they interact with charged particles from the solar wind and solar Extreme Ultra-Violet (EUV) photons.
These former events may result in a charge exchange and ionization of the helium atoms, removing them from the neutral population.
The interactions with EUV photons with the wavelength less than 50.4 nm result in photo-ionization, and with those within the 58.4 nm line result in so-called radiation pressure (generally neglected in calculations).
Radiation pressure is important for neutral hydrogen atoms, because depending on the solar activity phase, the average force generated by collisions between hydrogen and \lya{} photons can compensate the gravity force of the Sun.

Helium atoms have a greater mass than hydrogen and the solar spectral line responsible for helium ionization is much weaker than \lya{}. As a result, radiation pressure acting on neutral helium atoms is 2-3 orders of magnitude smaller than for hydrogen.
This is the main reason why it has always been neglected in all heliospheric models.

Both global and kinetic models of the heliosphere are currently very sophisticated \citep{pogorelov_etal:09a,heerikhuisen_etal:10a,katushkina_etal:14c, izmodenov_alexashov:15a,kornbleuth_etal:21b}.
They include more detailed descriptions of physical phenomena and even the most subtle effects can be comparable with model uncertainty. Therefore, we decided to calculate the extent of the uncertainty introduced in our model (WTPM) by neglecting the radiation pressure acting on neutral helium atoms.

In Section \ref{sec:theory}, we show how we calculate radiation pressure for helium atoms.
We use observational data from the SDO/EVE detector for absolute calibration of the solar spectral line HeI ($\lambda=58.4$ nm) and derive a formula for computing radiation pressure along the trajectory.
In Section \ref{sec:sim} we describe our simulations and provide the list of parameters that we use.
Section \ref{sec:results} shows an analysis of the trajectories of individual helium atoms (Subsection \ref{sec:atoms}), density of ISN He (Subsection \ref{sec:dens}), the IBEX-Lo flux (Subsection \ref{sec:IBEX}), and the IMAP-Lo flux (Subsection \ref{sec:IMAP}), calculated with and without radiation pressure.
Finally, in Section \ref{sec:summary}, we summarize and conclude our results.

\section{Radiation pressure - theory}
\label{sec:theory}

\subsection{Historical measurements of the HeI 58.4 solar spectral line}
The first measurements of the solar He I 58.4 nm line of helium were done in the 60's and 70's of the 20th century using rocket experiments \citep{hall_hinteregger:70a, cushman_etal:75a, doschek_etal:74a, maloy_etal:78a}.
Depending on the method that was used during the observation, the full width at half maximum (FWHM) and/or the total irradiance of the HeI line were measured.
Table \ref{tab:history} presents selected historical measurements.

The width of the solar emission line of HeI (58.4 nm) is not constantly monitored.
However, we have several measurements taken over the last few decades, which suggest that the width of the spectral line remain unchanged regardless of the phase of the solar cycle.
Its height, on the other hand, depends on the level of solar activity.
In this publication, we assume, following \citet{lallement_etal:04a}, that the width and the Gaussian shape of the line are constant over time, and normalization is based on the measurements from the Solar Dynamics Observatory (SDO\footnote{\url{https://www.nasa.gov/mission_pages/sdo/main/index.html}}).

\begin{deluxetable}{cccc}
\tablecaption{\label{tab:history} Selected historical measurements of the HeI line properties.}
\tablehead{\colhead{reference} & \colhead{obs date} & \colhead{FWHM [nm]} & \colhead{I$_{tot}$ [ph cm$^{-2}$ s$^{-1}$]}}
\startdata
\citet{hall_hinteregger:70a}                 & Mar 11 1967   &                     & $ 0.89 \times 10^9 $          \\                         
                                             & May 15 1967   &                     & $ 1.6 \times 0.89 \times 10^9 $\\
\citet{cushman_etal:75a}                     & Aug 30 1973   & $0.01$ (active)     & $ 1.3 \times 10^{9} \pm$ 40\% \\
                                             &               & $0.008$ (quiet)     &                               \\
\citet{doschek_etal:74a}                     & Sep 21 1973   & $0.014 \pm 0.0015$  &                               \\
\citet{maloy_etal:78a}                       & Nov 27 1974   & $0.0122 \pm 0.0010$ & $ (2.6 \pm 1.3) \times 10^9$  \\
\citet{phillips_etal:82a}                    & Nov 21 1974   &                     & $ 2.6 \times 10^9 \pm$ 50\%   \\
                                             & Sep 29 1977   & $0.0128 \pm 0.0020$ & $ 3.9 \times 10^9 \pm$ 50\%   \\
                                             & Sep 30 1980   & $0.0101 \pm 0.0010$ & $ 4.2 \times 10^9 \pm$ 50\%   \\
\citet{mcmullin_etal:04a,lallement_etal:04a} & 1996-2001     & 0.0092-0.0155       & $ 1.5  \times 10^9 $ \\
\citet{delZanna_andretta:15a}                & 1998-2015     &                     & (1.0 - 2.0) $\times 10^9$ \\
\enddata
\end{deluxetable}

\subsection{The SDO/EVE measurements of the HeI solar spectral line}
Since 2010 there are constant measurements available of the solar EUV spectrum taken by the SDO/EVE instrument \citep{woods_etal:12a} that is orbiting Earth.
The full resolution of the spectrum is 0.02 nm.
Apart from the spectrum, level 3 of the data products provides also a list of spectral lines with estimated total irradiance for each of them.
Among the listed lines there is the HeI line (58.4 nm) that we are interested in.
Currently, we are using Version 7 of the data that are available via the website \footnote{\url{https://lasp.colorado.edu/eve/data_access/evewebdataproducts/merged/}}.
The data are averaged over the Carrington rotation period (see the left panel in Figure \ref{fig:sdo}).
The ratio of the extreme values of the fluxes (maximum to minimum) for the most recent solar cycle is $\sim$1.4.

The level of the solar spectrum background has been estimated based on measurements around the wavelength $\lambda=$65 nm.
The ratio of the HeI line flux to the flux within 1 nm of 65 nm is around 32 and weakly dependent on the solar cycle phase.
The background level may be significant for atoms whose radial velocities are Doppler-shifted away from the center of the HeI line.
However, in our case, this will be a secondary effect compared to the already weak effect of radiation pressure.

\subsection{The HeI line model}
The definitions and units of the physical constants used in this paper are given in Table \ref{tab:units}.

\begin{deluxetable}{llll}
\tablecaption{\label{tab:units} Physical constants in cgs units. In our numerical simulations, we use more precise values based on the National Institute of Standards and Technology (NIST) standards.
}
\tablehead{\colhead{constant} & \colhead{symbol} & \colhead{value} & \colhead{unit}}
\startdata
elementary charge  & $e$   & $4.8032 \times 10^{-10}$ & $ \text{cm}^{3/2} \text{ g}^{1/2} \text{ s}^{-1}$\\
electron mass     & $m_\text{e}$ & $9.1094 \times 10^{-28}$ & g \\
speed of light    & $c$     & $2.9979 \times 10^{10}$  & $ \text{cm s}^{-1}$\\
He I wavelength & $\lambda_0$ & $58.433 $ & nm\\
astronomical unit & $r_\text{E}$ & $1.496 \times 10^{13}$ & cm\\
helium mass & $m_\text{He}$ & $6.6465 \times 10^{-24}$ & g\\
Gauss gravity constant $^a$ & $k^2=GM_\sun$ & $2.959122130672713\times 10^{-4}$ & au$^3$ day$^{-1}\, M_\sun$\\
spectral flux & $I_\lambda$ & time dependent & $\text{ph cm}^{-2} \text{s}^{-1} \text{ nm}^{-1}$\\
He oscillator strength $^b$ & $f_{\text{osc}}$ & $0.27625 $ & dimensionless\\
combination & $\frac{\pi e^2}{m_\text{e} c}$ & $0.0265402$ &$\text{cm}^{2} \text{ s}^{-1}$\\        
\enddata
\tablenotetext{a}{In the simulations, we use $k^2$ combination, because Gauss gravity constant is known with much greater accuracy than the $M_\sun$ and G separately, however in the equations we keep $GM_\sun$ for clarity.}
\tablenotetext{b}{\cite{drake:06a}}
\end{deluxetable}

We assume that the HeI spectral line shape is given by the Gauss function normalized in such a way that the value in the centre of the line is equal to $I_0=I_\lambda(\lambda_0)$, with FWHM=0.0136 nm (following \citet{lallement_etal:04a} model):

\begin{equation}
I_{\lambda}(\lambda)=I_0 \exp{\left[ -\frac{1}{2}\left( \frac{\lambda - \lambda_{0}}{\sigma} \right)^2 \right] },
\label{eq:gauss}
\end{equation}

Gauss function described by Equation \ref{eq:gauss} takes the value of $\frac{1}{2}I_0$ for $\lambda_h= \pm 2\sqrt{2\log{2}}$.
From the definition FWHM$=2\lambda_h$, thus standard deviation $\sigma$ can be expressed by FWHM as:
\begin{equation}
\sigma=\frac{FWHM}{2\sqrt{2 \log{2}}}.
\end{equation}

The equivalent width ($w$) results from the comparison of the integral of the Gaussian function and the area of the rectangle with the base $w$ and height equal to the $I_0$:

\begin{eqnarray}
I_{tot}=\int_{-\infty}^{\infty} I_{\lambda}(\lambda) d\lambda = I_0 \sqrt{2\pi} \sigma = w I_0\\
w=\sqrt{2\pi} \sigma = \sqrt{2\pi} \frac{FWHM}{2\sqrt{2 \log{2}}}= 1.064  FWHM
\end{eqnarray}

The total flux $I_{tot}$ in the HeI line is measured by the SDO/EVE instrument.
Assuming the Gaussian shape and equivalent width $w$, we can calculate the flux in the centre of the HeI line as:

\begin{equation}
I_0=\frac{I_{tot}}{\sqrt{2\pi} \frac{FWHM}{2\sqrt{2\log{2}}}}
\label{eq:I0}
\end{equation}

\subsection{Radiation pressure}
Spectral irradiance in the HeI line can be expressed as a ratio between solar gravity force and the force caused by the radiation pressure.
A detailed derivation of the radiation pressure formula is given in Appendix A of \citet{kubiak_etal:21a}.
Here we give a definition of $\mu$ and a scaling factor that can be used to convert the irradiance into the $\mu$ factor (Equation A17 from the aforementioned paper adapted to helium atoms) in the atom's rest frame:

\begin{align}
\label{eq:radpress}
\nonumber
\mu&=\frac{\text{P}_\text{rad}}{|\text{F}_\text{g}|}\\ \nonumber
&=I_0 \frac{\pi e^2}{m_e\, c} \frac{h \lambda_0}{c} \,f_{\text{osc}} \frac{r_\text{E}^2}{GM_\sun m_\text{He}}\\
&=\frac{I_0}{p_\text{He}},\\
\label{eq:ph}
p_\text{He} & = \left[\frac{\pi e^2}{m_e\, c} \frac{h \lambda_0}{c} \,f_{\text{osc}}\, \frac{r_\text{E}^2}{GM_\sun m_\text{He}}\right]^{-1} = 4.162472674170337\times 10^{13} & \text{ ph s}^{-1}\,\text{cm}^{-2}\,\text{nm}^{-1}
\end{align}

In the Sun's rest frame, we must take into account the radial component of the atom's velocity, which, due to the Doppler effect, causes photons of slightly different wavelengths to be absorbed.
The radial velocity can be easily related to the shift in wavelength ($\Delta \lambda$) using the following formula:

\begin{equation}
v_r(\lambda)=-\frac{\Delta \lambda}{\lambda_0}c
\label{eq:vr}
\end{equation}

In our calculations, we use the radiation pressure dependent on the instantaneous radial velocity calculated at each point of it's trajectory and normalized by the rescaled total solar irradiance, that is time dependent.
Thus, the formula we use to calculate the $\mu$ factor is as follows:

\begin{align}
\label{eq:radpress_calc}
\nonumber
\mu(t,\Delta \lambda(v_r))&=\frac{I_{\lambda}(t,\Delta \lambda(v_r))}{p_\text{He}}\\
I_{\lambda}(t,\Delta \lambda(v_r)) & = \frac{I_{tot}(t)}{\sqrt{2\pi} \frac{FWHM}{2\sqrt{2 \log{2}}}} \exp{\left[ -\frac{1}{2}\left( \frac{\Delta \lambda(v_r)}{\frac{FWHM}{2\sqrt{2 \log{2}}}} \right)^2 \right] }
\end{align}

We neglect two factors in our analysis: the absorption effect and the latitudinal structure of solar radiation in the HeI line. We consider these factors to be of secondary importance compared to the influence of radiation pressure, which is the main focus of this article.

\begin{figure*}[ht!]
\centering
\includegraphics[width=1\linewidth]{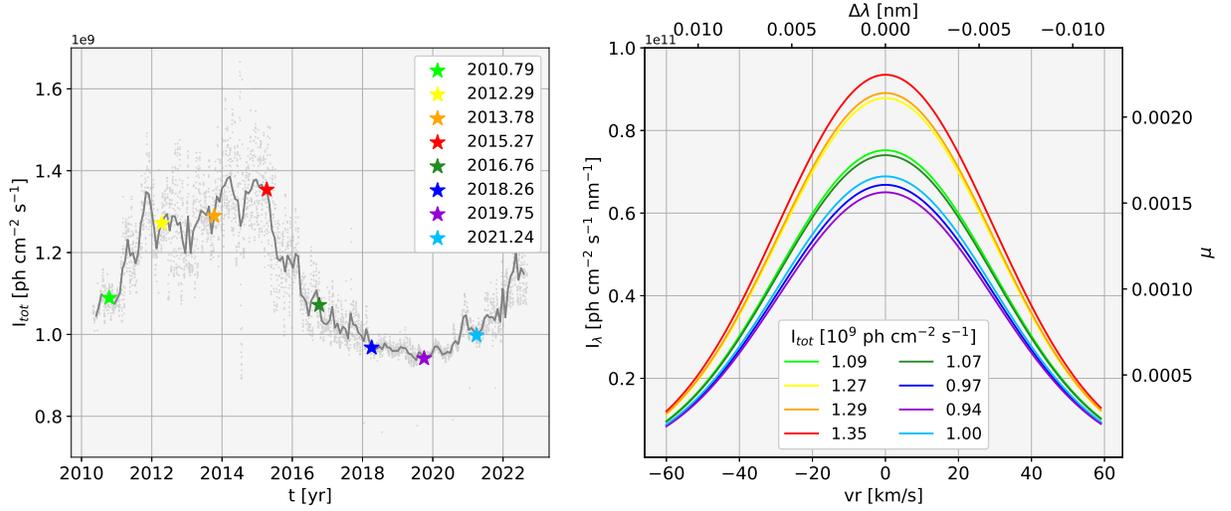}
\caption{Total flux in the HeI solar line provided by the SDO/EVE instrument (left panel) and example profiles generated using our model described in the text and by Equation \ref{eq:radpress_calc} (right panel). The colors represent different phases of solar activity and correspond to each other on both panels.}
\label{fig:sdo}
\end{figure*}

The right panel in Figure \ref{fig:sdo} shows a few example shapes of the HeI profile line calculated using our model.
There are two vertical axis labels: the left one shows the value of the spectral irradiance in physical units (ph s$^{-1}$ cm$^{-2}$ nm$^{-1}$), while the right one shows the value of the $\mu$ factor that is essential to estimate the radiation pressure strength.
In our models, we use radiation profiles as a function of the radial velocity of the interacting helium atom, but for better comparison with the values given in Table \ref{tab:history} we present both horizontal axes (the lower shows radial velocity, while the upper shows the corresponding wavelength calculated from the center of the line).
The total irradiance for those profiles was taken from the SDO/EVE measurements shown on the left panel as color stars.
The colors of the stars in the left panel correspond to the colors of the lines in the right panel.
Gray dots on the left panel are the daily SDO/EVE measurements, while the gray solid line shows the average over the Carrington rotation period.
We choose stars to represent different levels of solar activity during the last cycle.

Even in the high activity period, when the HeI flux is strong (the red line), radiation pressure is at the level of $2.2 \times 10^{-3}$, which means that the force caused by radiation is only 0.22\% of the gravitational force acting on the helium atom.
At the same time, radiation pressure acting on the H atoms is of the order of 1 and can compensate or even overcome the Sun's gravity force.
Therefore, we do not expect radiation pressure to affect He trajectories significantly.

\section{Simulations}
\label{sec:sim}

For simulations that take into account the influence of radiation pressure, we use the WTPM model described by \citet{tarnopolski_bzowski:08a}, which has been adapted to follow the behaviour of the helium atoms instead of hydrogen.
This is the hot model of the interstellar gas distribution, and individual atoms are tracked numerically by solving the equations of motion, taking into account the radiation pressure dependent on the radial velocity of the atom.
The time it takes for atoms to reach the vicinity of the Earth's orbit differs depending on the population \citep{bzowski_etal:19a}.
The primary population is faster, so the arrival time is much shorter (as seen in panels (a) and (c) of Figure \ref{fig:orbit}).
In our simulations, variable radiation pressure is turned on in 2010 when direct measurements from SDO/EVE become available.
Earlier stages of the atom trajectory are calculated assuming a constant radiation pressure of $\mu=1.4\times 10^{-3}$. Even the slow secondary population observed in the detector in 2015 was at a distance greater than 10 au in 2010.
This means that the total force indicating radiation pressure, which decreases with the square of the distance, is 100 times smaller there than that at a distance of 1 au.
Hence, at 10 au the total force is anyway very small, and a small modification of solar gravity force by the radiation pressure force does not change the trajectories significantly.
Panels (b) and (d) of Figure \ref{fig:orbit} show that the maximum radiation pressure value for the considered atoms is only $\mu=2.2\times 10^{-3}$.
Therefore, the simplification made in our simulations is justified.

Simulations without radiation pressure, which we use in comparative analysis, are done by setting radiation pressure factor $\mu=0$.

In both cases, we used a model consisting of two populations: a primary population corresponding to pure interstellar gas, and a secondary population known as the "Warm Breeze" \citep{kubiak_etal:14a, kubiak_etal:16a, bzowski_etal:17a, bzowski_etal:19a}, which is a result of charge exchange interactions in the outer heliosheath, but in our approach is introduced as a separate population, non-interacting with the other one.

The distribution of both populations at the boundary of our calculations is described by the Maxwell-Boltzmann distributions placed at a distance of 150 au with the parameters listed in Table \ref{tab:pop_params}.
In this particular case, we can neglect the charge exchange at a distance of no more than 150 au from us, because we already introduced the secondary population. If we want the secondary population to synthesize from the primary population, then we do need to allow for charge exchange processes, but we also need to extend the computational boundary to a distance of about 1000 au. This was done in the \cite{bzowski_etal:17a} paper. From the point of view of this paper, it does not matter which assumption we use, as long as we do it consistently for both cases: with and without radiation pressure.
\cite{mccomas_etal:15b} showed that beyond 150 au ISN He parameters are not changing significantly, therefore we can assume that the calculation boundary at 150 au is as good as in the infinity.
The charge exchange ionization of both those populations in the inner heliosheath is potentially important for the ENA (Energetic Neutral Atoms) production, but this reaction modifies the two populations negligibly in comparison with anyway small modification due to photoionization in this region.

\begin{deluxetable}{ccccccc}[!h]
\tablecaption{\label{tab:pop_params} Parameters of the two He populations used in this paper.}
\tablehead{\colhead{population} & \colhead{flow lon [$\degree$]} & \colhead{flow lat [$\degree$]} & \colhead{v [km s$^{-1}$}] & \colhead{T [K]} & \colhead{n [cm$^{-3}$]} & \colhead{ref}}
\startdata
primary    & 255.745 & 5.169 & 25.784 & 7443 & 0.0141911$^a$  & \citet{bzowski_etal:15a}\\
secondary  & 251.57  & 11.95 & 11.28  & 9480 & 0.000808893 & \citet{kubiak_etal:16a} \\  
\enddata
\tablenotetext{a}{\cite{gloeckler_geiss:04a} obtained the total He density at 0.015 cm$^{-3}$. This is distributed between the primary and secondary populations as listed. }
\end{deluxetable}

We calculated ISN He density distribution in the heliosphere, the IBEX-Lo signal, and the IMAP-Lo signal in order to conduct in-depth research on the impact of radiation pressure on helium.

\subsection{The ISN He density}
Simulations of the density distribution of helium in the heliosphere were conducted for the minimum solar activity in 2019 and for the maximum solar activity in 2015.
The computational grid covered three planes: the ecliptic plane, the polar plane passing through the upwind and downwind directions, and the polar plane passing through the crosswind directions.
The nodes were spaced at 5-degree intervals in the azimuthal angle.
The distances from the Sun ranging from 0.1 to 54 AU were equally spaced on a logarithmic scale.

\subsection{The IBEX-Lo signal}
In-situ measurements of ISN He are done by the Interstellar Boundary Explorer (IBEX) \citep{mccomas_etal:09a}. 
The ISN atoms are observed using the IBEX-Lo instrument \citep{fuselier_etal:09b}, which is a time-of-flight mass spectrometer.
It was designed to detect energetic neutral atoms in Earth's orbit, and measure the direction from which they came and the time of arrival.
The line of sight of the IBEX-Lo detector is inclined to the spacecraft spin axis at an angle of 90$\degree$ and scans a strip along a great circle with every spin.
The spin axis is pointing towards the Sun and its position changes once per week.
Our simulations take into account the gas distribution, geometry of the observations, velocity of the spacecraft, and collimator effects \citep{sokol_etal:15a}.
Following \citet{kubiak_etal:16a, bzowski_etal:19a}, we choose spin angles for each orbit within 204-306$\degree$ range, to focus on the peak of the helium signal.

\subsection{The IMAP-Lo signal}
IMAP-Lo is one of the 10 instruments of the Interstellar Mapping and Acceleration Probe (IMAP) mission \citep{mccomas_etal:18b} that will be launched in early 2025.
IMAP-Lo will be mounted on a pivot platform that will allow to change the elongation angle (an angle between the rotation axis of the spacecraft and the detector's axis) during the mission within range 60-160$\degree$.
The detector's line of sight will be adjusted to observe the most interesting signals depending on the actual spacecraft position.
IMAP will be placed in a stable orbit around the Sun-Earth L1 Lagrange point.
The spin axis will be pointing 4$\degree$ behind the Sun in the ecliptic and adjusted daily to maintain the designed position.
The expected sensitivity of the instrument is better than that for IBEX-Lo.
Longer exposure time on IMAP-Lo will assure much larger count numbers, and as a consequence, statistical errors will be smaller than that on IBEX-Lo in the same regions of observation.
Our simulations cover a broad range of possible observation configurations.
The elongation angle is sampled from 48-180$\degree$ with a cadence of 4$\degree$, with an additional node for 90$\degree$.
The longitude of the spacecraft, that can be translated into time expressed by days of the year, is sampled every 4 days.
We use a full range of the spin angles within each geometry of observation, sampled every 6$\degree$.

\section{Results}
\label{sec:results}

\subsection{Selected He atoms trajectories}
\label{sec:atoms}
Using the WTPM code, we can track the trajectories of individual atoms depending on their measured energy, detector location, ionization rate, and solar activity phase.
We have analysed trajectories of five atoms detected during solar minimum and five atoms detected during solar maximum.
We selected a few types of atoms:
\begin{itemize}
\item typical atoms from the primary population, where the spin angles correspond to the IBEX-Lo peak of the simulated flux during one orbit (pr orbit 436b and pr orbit 275b)
\item atoms from the primary population, where the spin angles correspond to the small IBEX-Lo flux, but the radiation pressure effect is relatively strong (pr orbit 438b and pr orbit 278a)
\item atoms from the secondary population, where the simulated IBEX-Lo flux is small, but the radiation pressure effect is relatively strong (sc orbit 438b and sc orbit 278a)
\item atom from the secondary population detected by the IMAP-Lo detector for DOY=265 and elongation angle $\epsilon_{FOV}=60 \degree$, where the simulated IMAP-Lo flux is high, but radiation pressure effect is small (sc $\epsilon_{FOV}=60\degree$).
\item atom from the secondary population detected by the IMAP-Lo detector for DOY=265 and elongation angle $\epsilon_{FOV}=90 \degree$, where the simulated IMAP-Lo flux is small, but radiation pressure effect is relatively strong (sc $\epsilon_{FOV}=90\degree$).
\end{itemize}

\begin{figure*}[ht!]
\centering
\includegraphics[width=1\linewidth]{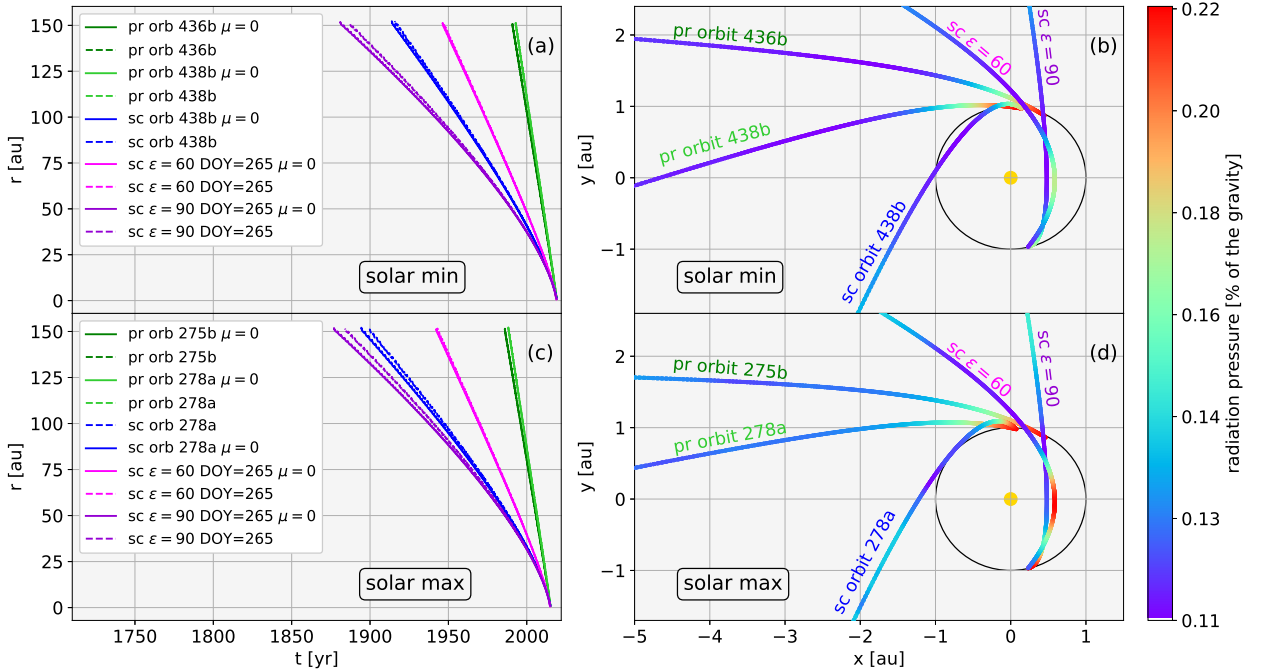}
\caption{Projection of the trajectories of selected He atoms on ecliptic plane during solar minimum (panels (a) and (b)) and solar maximum conditions (panels (c) and (d)).
The left panels ((a) and (c)) present radial distance from the Sun as a function of time.
Solid lines show calculations without radiation pressure, while dashed lines show trajectories with radiation pressure included.
The right side panels ((b) and (d)) show the magnitude of the radiation pressure along the trajectory in the heliographic coordinate system.
The gold circle in the centre shows Sun's position and the black circle reflects Earth's orbit.
}
\label{fig:orbit}
\end{figure*}

Results are shown in Figure \ref{fig:orbit}.
The left side panels show how the radial distance of the atom from the Sun is changing over time.
Solid lines correspond to the calculations without radiation pressure, and dashed lines show how the distance would change if we include radiation pressure.
The difference between the solid and dashed lines is visible only far away from the Sun (more than 50 au) for the secondary population detected by IBEX-Lo (blue lines) and for the secondary population detected by IMAP-Lo with the elongation angle $\epsilon_{FOV}=90\degree$ (dark violet lines).
Those atoms travel slower through the heliosphere, therefore their trajectories are more affected by the radiation pressure than those from the primary population.
In the case of the primary population atoms (green lines) or even atoms from the secondary population detected by IMAP-Lo with elongation angle $\epsilon_{FOV}=60\degree$ (magenta lines), the effect of radiation pressure is negligible.

The right panels of the same Figure show the strength of the radiation pressure expressed as a percentage of the Sun's gravity force that is acting on the atoms in the Cartesian heliographic coordinate frame.
It means that the radiation pressure is ~0.22\% of the Sun's gravity in the red parts of the trajectories and is below 0.11\% for dark blue parts of the trajectories.
The trajectories end at the detector position (IBEX or IMAP depending on the atom).
The strongest radiation pressure is seen close to the Sun, but the exposure time is relatively short, so the overall effect on the trajectory is small.
The trajectory marked as $\epsilon_{FOV}=90$ looks like it is getting closer to the Sun than that for $\epsilon_{FOV}=60$, so radiation pressure should be bigger there.
But, in fact, it's just an effect of projection on the xy plane, the $\epsilon_{FOV}=90$ atom is much farther from the Sun in the direction of the z axis, hence the small magnitude of the radiation pressure.
By comparing panel (b) and (d), we can see the time dependence caused by the solar cycle.
Radiation pressure as a whole is higher during solar maximum (panel (d)), because the solar line HeI is stronger at that time.
But we don't observe any important differences between the trajectories detected during solar maximum and solar minimum.
For all selected atoms even the strongest effect is still much below 1\% of the gravitational force.

\newpage
\subsection{The ISN He density}
\label{sec:dens}
Using the WTPM numerical code we obtained the distribution of He neutral atoms in the Heliosphere.
We compared density distributions calculated with and without radiation pressure for low and high solar activity (2019 and 2015, respectively).
Results show that the difference in density is smaller than 0.2\% regardless of the solar cycle phase.
The biggest changes can be seen in the case of atoms that have passed close to the Sun, as they were exposed to the strongest radiation pressure.
That is why the downwind direction was slightly more affected due to the atom's longer exposure to the effect of radiation pressure.
However, in both cases, direct observation of these atoms will be impossible.
Therefore, we can conclude that radiation pressure neither affects studies of the pick-up ions (PUIs) nor the ionization rates estimates.

\subsection{The IBEX-Lo signal}
\label{sec:IBEX}
\begin{figure*}[ht!]
\centering
\includegraphics[width=1\linewidth]{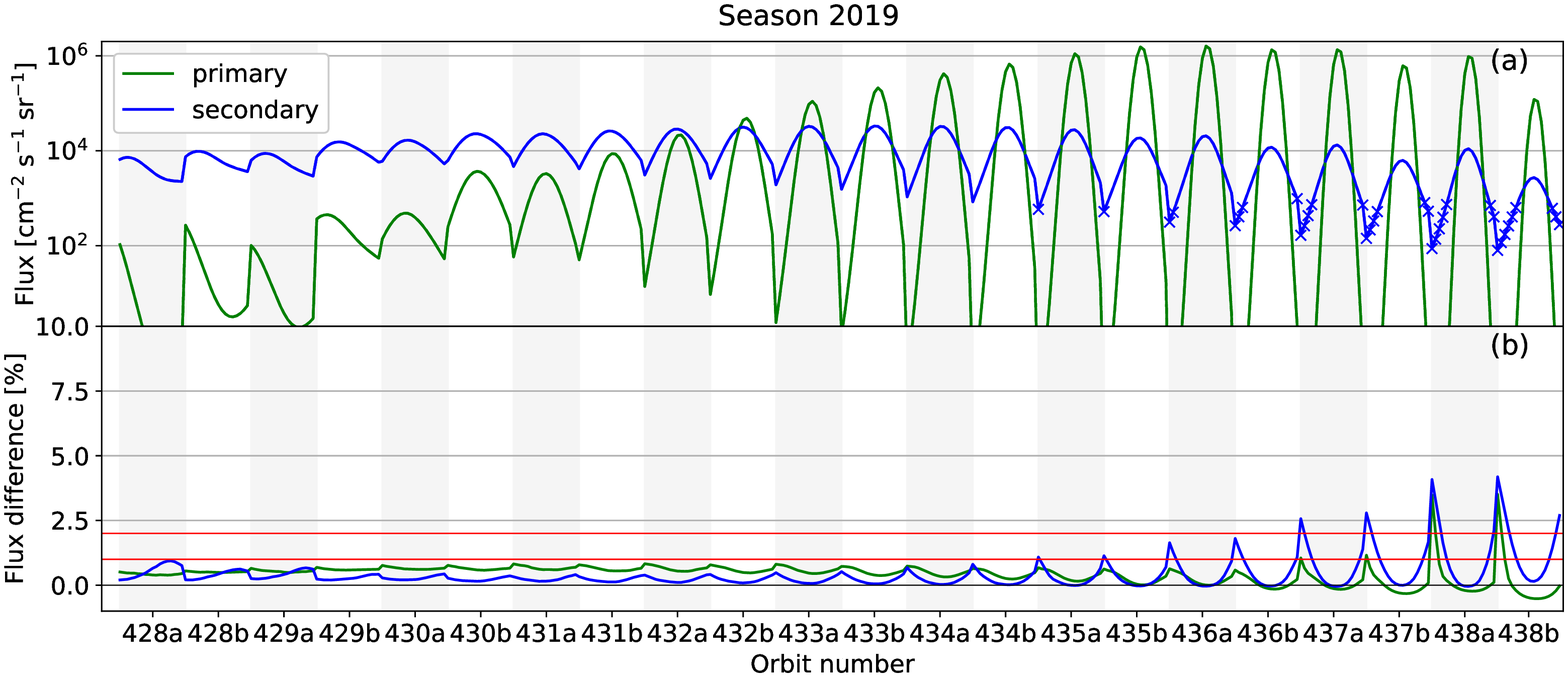}
\includegraphics[width=1\linewidth]{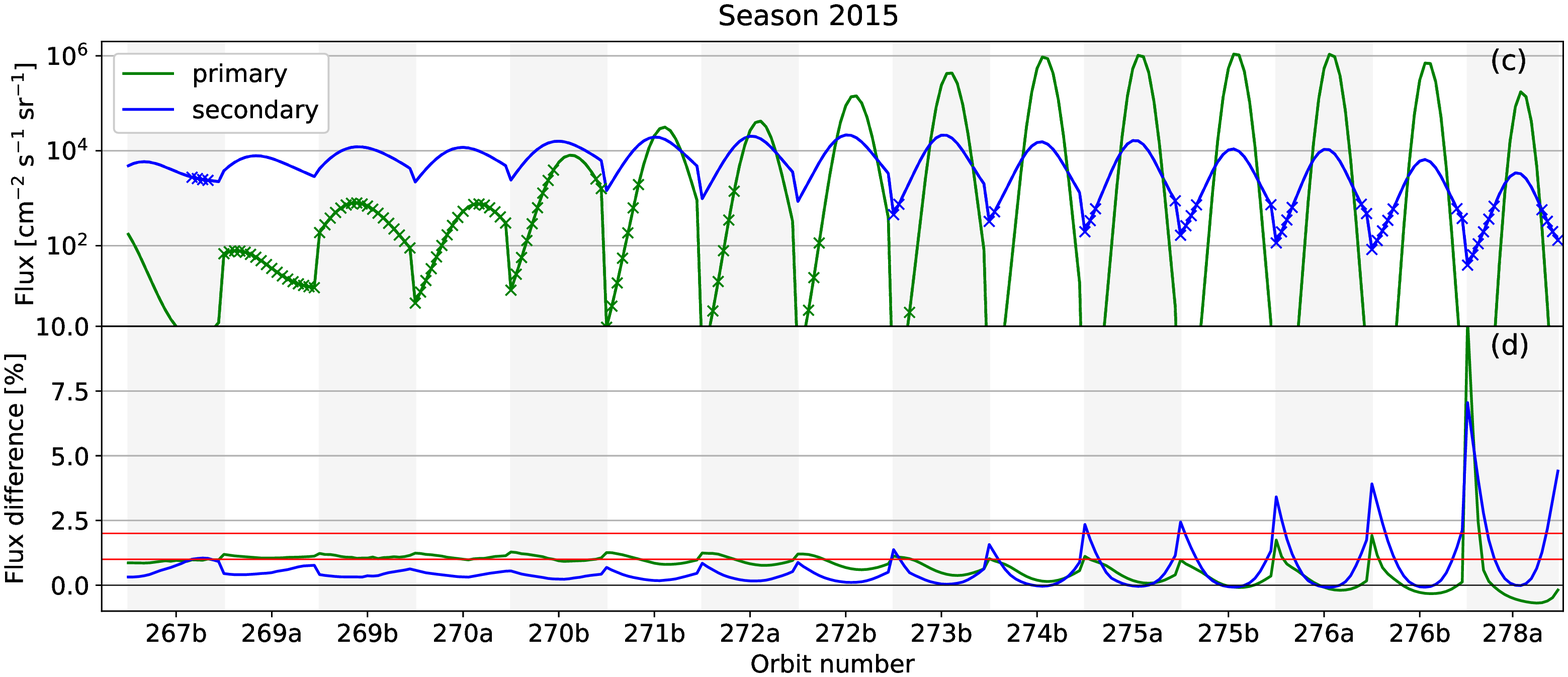}
\caption{The IBEX-Lo simulated flux for low solar activity in season 2019 (upper plot) and high solar activity in season 2015 (lower plot). Panels (a) and (c) show the  simulated flux for the helium primary population (green line) and the secondary population (blue line). Panels (b) and (d) show the relative difference between the flux calculated with radiation pressure and without it. The "x" symbols on panels (a) and (c) mark the points for each IBEX orbit where the flux difference shown in panels (b) and (d) is greater than 1\%. Red horizontal lines mark the flux difference at the level of 1\% and 2\%.}
\label{fig:ibex_flux}
\end{figure*}

The flux of atoms measured by IBEX-Lo is sensitive to the ionization rate, and - for hydrogen - the radiation pressure.
It was shown that the details of the radiation pressure model for hydrogen atoms can change the estimated flux significantly \citep{tarnopolski_bzowski:09, IKL:18b}.
One of the main goals of this work is to investigate the sensitivity of the observed signal to the radiation pressure for helium.
We analysed simulations of 12 observational seasons of IBEX from 2009 to 2020.
This time span covers more than a full cycle of solar activity.
Figure \ref{fig:ibex_flux} shows two observational seasons that are representative of low (season 2019) and high (season 2015) solar activity.
The upper panels in each season ((a) and (c)) show the simulated IBEX-Lo flux of the primary (green line) and the secondary (blue line) populations of the ISN He atoms.
The lower panels ((b) and (d)) show relative differences between the fluxes calculated with and without radiation pressure.
The red horizontal lines mark levels of 1\% and 2\%, and the "x" symbols on the flux plots mark the points where the difference is above 1\%.

The strongest effect can be seen when the flux itself is very small.
The impact of radiation pressure on the primary population flux is visible in places where it is obscured by a much stronger secondary population flux.
The only area where radiation pressure plays a significant role is in the wings of the "Warm Breeze" in the second half of the observing season.
\citet{bzowski_etal:19a} showed that in this region the differences between the two "Warm Breeze" treatment (the primary and the secondary population as a separate distributions or the primary population and interactions that create the secondary component) appear, but the errors are too big to distinguish between those two phenomena (see Figure 4 in \cite{bzowski_etal:19a} or Figures 5 and 6 in \cite{kubiak_etal:16a}).

\begin{figure*}[ht!]
\centering
\includegraphics[width=1\linewidth]{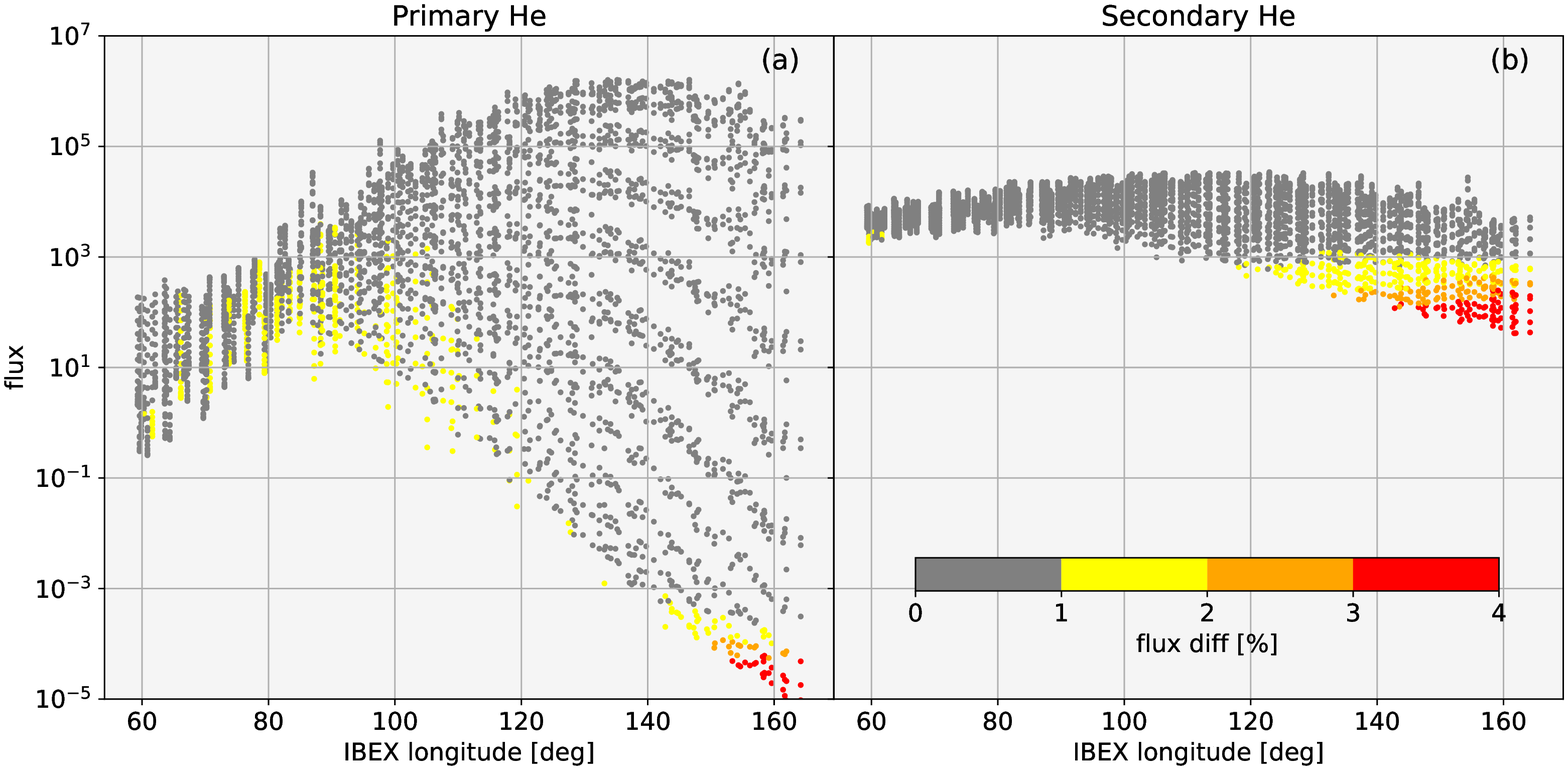}
\caption{Difference between the IBEX-Lo flux calculated with and without radiation pressure as a function of the flux and spacecraft ecliptic longitude for the primary (panel (a)) and the secondary (panel (b)) populations. The color scale displays the percentage difference in the fluxes.}
\label{fig:ibex_scatter}
\end{figure*}

Figure \ref{fig:ibex_scatter} shows the relationship between the difference in fluxes calculated with and without radiation pressure as a function of the flux and the spacecraft's ecliptic longitude for the primary (panel (a)) and the secondary (panel (b)) populations.
The color scale allows for identification of observation directions where this effect may be significant.
In both cases, the strongest effect is observed when the spacecraft is between 140 and 160 degrees ecliptic longitude.
For the primary population, a flux difference above 2\% occurs for very small flux values (below $10^{-3}$), so it can be neglected.
The secondary population is more susceptible to the effects of radiation pressure.
A flux difference above 2\% mostly occurs when the particle flux exceeds 100, which could potentially be significant.
But all those potentially significant points are located at the wings of the flux versus spin angle plots for each orbit (see blue "x" symbols in Figure \ref{fig:ibex_flux}), where the IBEX-Lo measurement errors are the largest (see Figure 4 in \cite{bzowski_etal:19a} or Figures 5 and 6 in \cite{kubiak_etal:16a} for reference).
Therefore, the effect of radiation pressure is below the IBEX-Lo sensitivity and there seems to be no chance to detect it.

\subsection{The IMAP-Lo signal}
\label{sec:IMAP}
\begin{figure*}[ht!]
\centering
\includegraphics[width=1\linewidth]{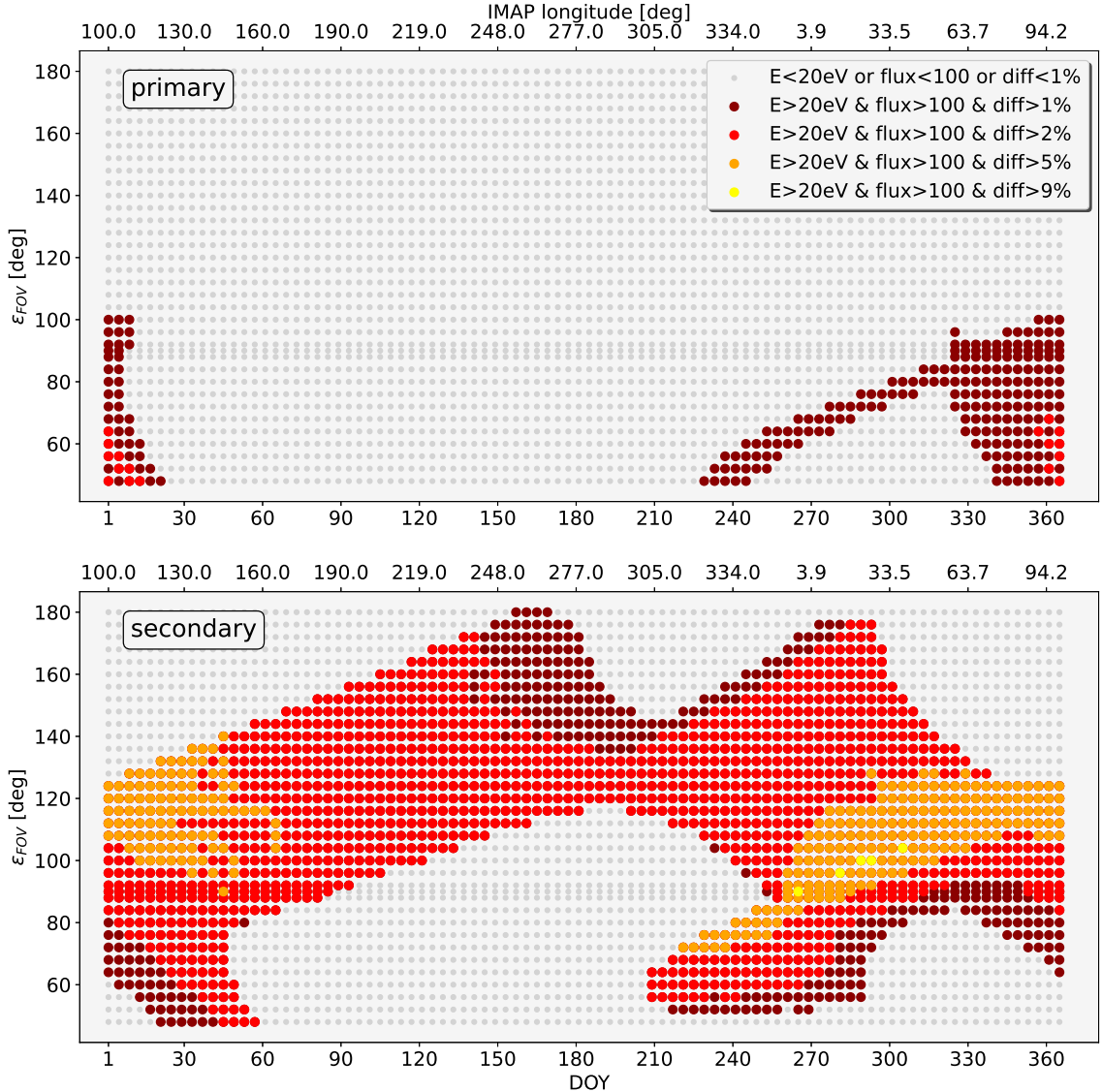}
\caption{Radiation pressure effect on the IMAP-Lo flux during high solar activity (as it was in 2015) as a function of the time and the elongation angle $\epsilon_{FOV}$. Colour points show where the difference between calculations with and without radiation pressure is greater than 1\% (dark red), 2\% (red), 5\% (orange), and 9\% (yellow), while the flux is above 100 cm$^{-2}$ s$^{-1}$ sr$^{-1}$ and the energy of the atoms is above 20 eV. The upper panel presents the primary population, while the lower one - the secondary population ("Warm Breeze"). The lower horizontal axis is scaled in days of the year, while the upper horizontal axis shows corresponding ecliptic longitudes of the spacecraft.}
\label{fig:imap_flux_diff_2D}
\end{figure*}

The design of IMAP-Lo will be similar to that of IBEX-Lo, but with the addition of a pivot platform that will provide new observational opportunities.
Details can be found in \citet{sokol_etal:19c} and Kubiak et al. 2023 (in preparation).
Figure \ref{fig:imap_flux_diff_2D} shows where we expect radiation pressure to be important in the time and pivot platform elongation angle parameter space.
The calculations that we show here assume high solar activity as it was during solar maximum in 2015.
IMAP will be launched during similar conditions in 2025.
During lower solar activity radiation pressure is weaker, therefore the considered effect is smaller.
Colour points show where radiation pressure is relevant, that is, the following conditions for at least one spin angle bin are fulfilled: the expected flux is above 100 cm$^{-2}$ s$^{-1}$ sr$^{-1}$, the energy of the atoms is greater than 20 eV, and the difference between the calculations with and without radiation pressure is above 1\% (dark red), 2\% (red), 5\% (orange), or 9\% (yellow).
Gray mark the spots of all grid points.
The primary population, shown in the top panel, is almost not affected.
The only regions where radiation pressure could be relevant are in the downwind direction (IMAP longitude $\sim 75 \degree$) for a low elongation angle of the pivot platform.
For the secondary population ("Warm Breeze"), a large part of the available observation configurations will be affected.

For both populations, especially affected will be a region where we expect an indirect beam of the He atoms (DOY$>$250).
This effect is understood given that an indirect beam is formed by the atoms that passed their perihelia, and hence were exposed to a stronger radiation pressure close to the Sun for a longer time than atoms from the direct beam.
This region can be compared to the IBEX-Lo "Warm Breeze" orbits.
The flux is comparable, so statistical errors of IMAP-Lo measurements shouldn't be larger than on IBEX-Lo (which are shown in Figure 4 in \cite{bzowski_etal:19a} or Figure 5 and 6 in \cite{kubiak_etal:16a}).
It means that radiation pressure effect causing a flux difference up to 9\% will likely be statistically significant.

\begin{figure*}[ht!]
\centering
\includegraphics[width=1\linewidth]{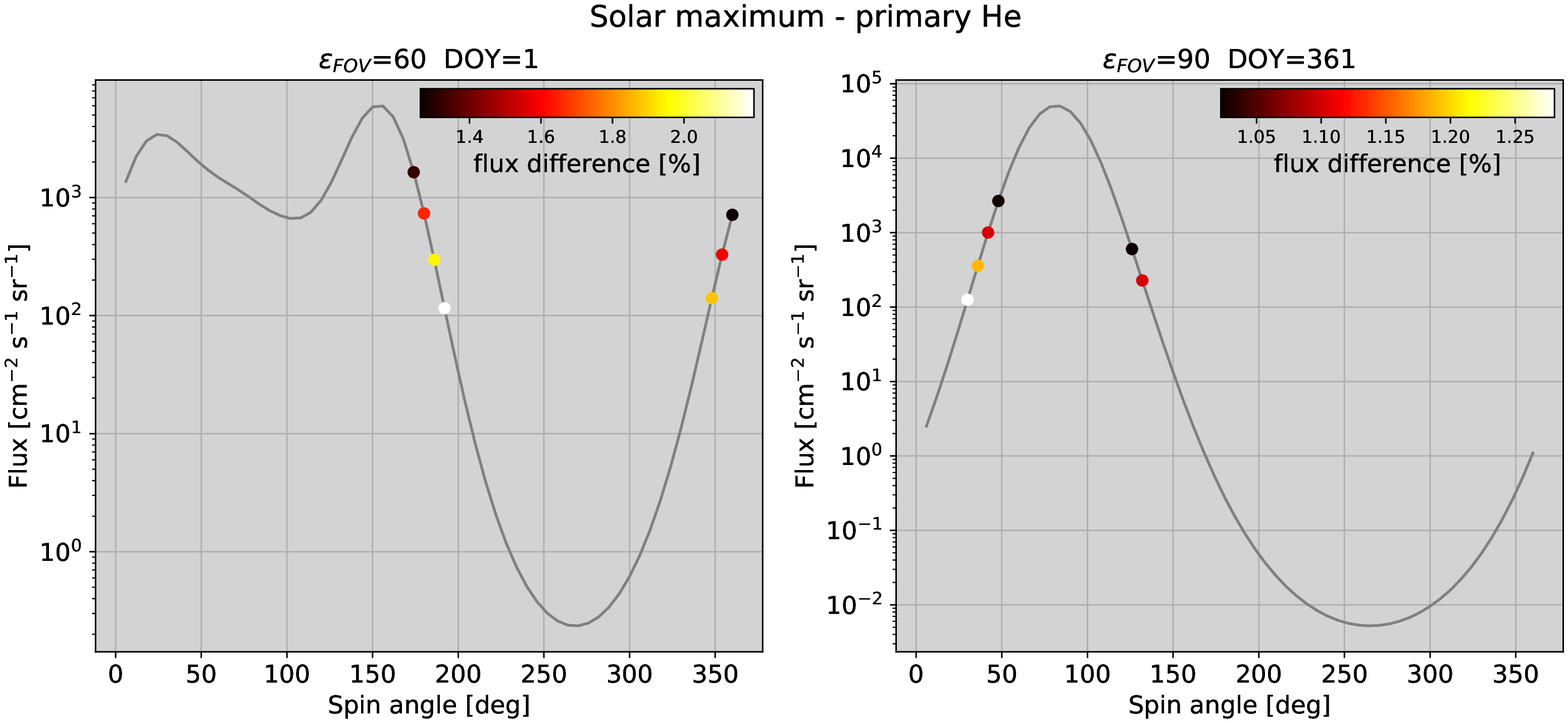}
\includegraphics[width=1\linewidth]{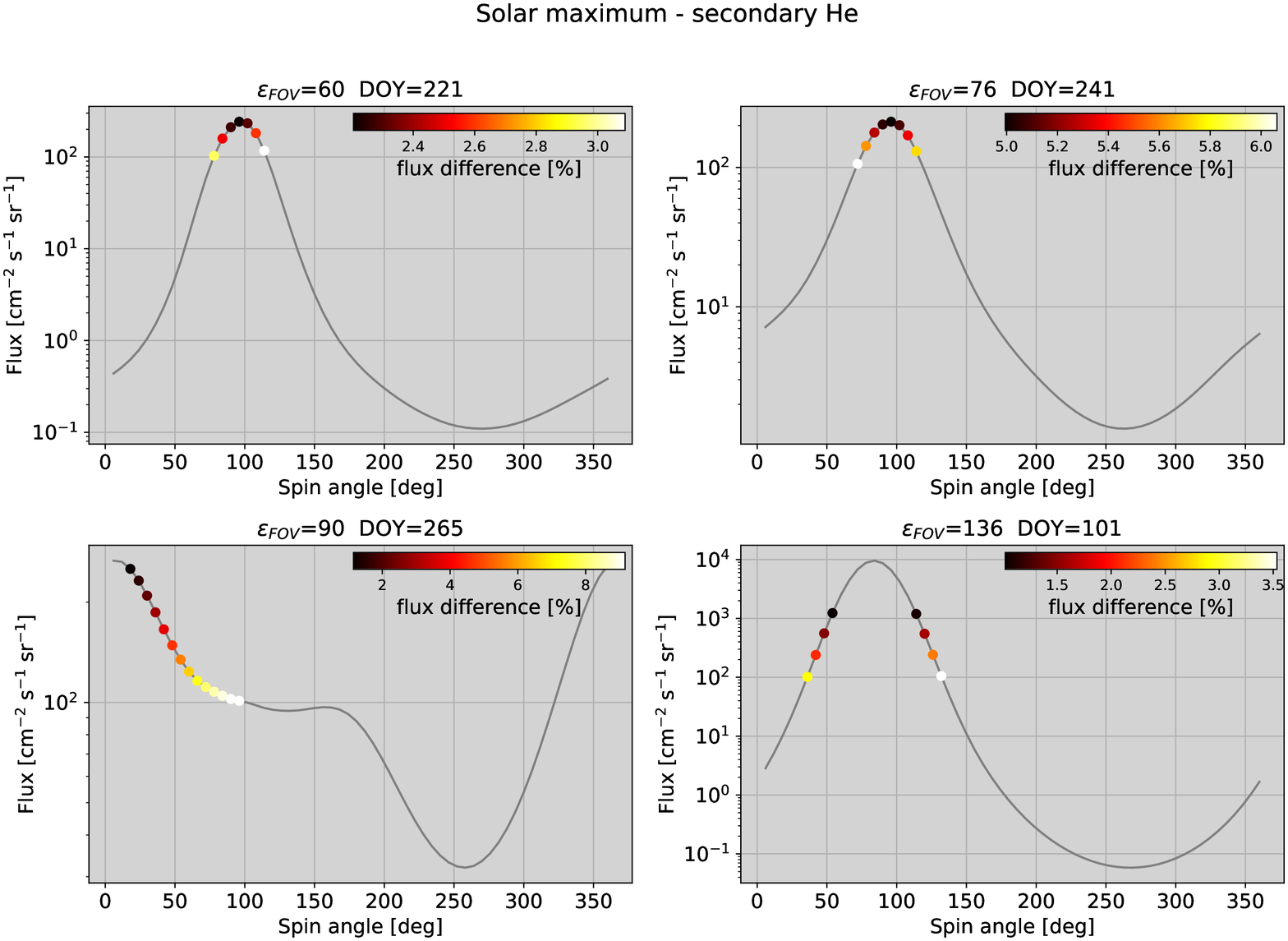}
\caption{The simulated IMAP-Lo flux of the primary and the secondary He as a function of the spin angle for a selected time of observation (in DOY) and elongation angle. Gray solid line shows the flux for a full range of angles, while points show only those pixels where the difference in the flux is greater than 1\%, the flux is above 100 cm$^{-2}$ s$^{-1}$ sr$^{-1}$, and the energy of the atoms is above 20 eV. The coloured scale shows the magnitude of the radiation pressure effect.}
\label{fig:imap_flux_diff}
\end{figure*}

For a given elongation angle ($\epsilon_{FOV}$) and day of the year (DOY), the simulated flux is changing with the spin angle of the spacecraft.
Selected examples are shown in Figure \ref{fig:imap_flux_diff} for the primary and the secondary He populations.
The elongation angles were taken as it is planned for IMAP-Lo observations.
Colour points mark those spin angles where all conditions are met (the flux above 100 cm$^{-2}$ s$^{-1}$ sr$^{-1}$, the energy above 20 eV, and the flux difference greater than 1\%).
The part of the orbit where the spin angles are smaller than 180$\degree$ is where the detected atoms are hitting the detector almost head-on.
Since the velocity of the spacecraft adds to the velocity of the atom, we will observe higher energies and the atom will be easier detected.
The other half of the orbit is less preferable due to much smaller energies (the atoms are chasing the spacecraft).

\newpage
\section{Summary and Conclusions}
\label{sec:summary}
We conducted a comparative analysis of the significance of the influence of radiation pressure on the density distribution of ISN He inside the Termination Shock (TS), as well as on the IBEX-Lo and IMAP-Lo signals.
We performed calculations for epochs of low and high solar activity.
The phase of the solar cycle did not significantly affect the obtained results, although during solar maximum the effect was slightly greater.
In the neutral helium density study, we found no significant effect of radiation pressure.
The densities obtained taking into account the radiation pressure were in agreement with the previous calculations (without radiation pressure) with an accuracy of more than 0.2\%.
Therefore, studies of the pick-up ions (PUIs) and ionization rate estimates will not be affected by the radiation pressure effect.
The secondary population flux seen by IBEX-Lo is sensitive to radiation pressure in late orbits outside the He flux maximum.
The radiation pressure effect in this region is 1-4\%, which is imperceptible given the current measurement errors.
The primary population seen by IBEX-Lo remains unchanged within the measurement errors.
Even though the effect of radiation pressure is relatively small, it affects similar regions of the signal with similar magnitude as other small effects such as: discrepancy between different models of the "Warm Breeze" and observations \citep{bzowski_etal:17a}, significance of the elastic collisions in the outer heliosheath \citep{swaczyna_etal:19a}, precise estimations of the flow direction \citep{swaczyna_etal:22b}.

The simulated neutral He flux expected in the IMAP-Lo observations will be affected by radiation pressure in a wide range of possible observation configurations.
A particularly large effect (above 9\%) will be seen in the indirect beam of the secondary He population.
Slightly smaller changes in the flux (at the level of 2-5\%) are visible in the vicinity of the indirect beam of the primary He population.
For a deeper analysis of the IMAP-Lo observations, it will sometimes be necessary to take into account the influence of radiation pressure on helium.

\acknowledgments 
The authors would like to thank Mike Gruntman for sharing his research on the topic and  an anonymous reviewer for help in improving this manuscript.
This study was supported by Polish National Science Center grants 2019/35/B/ST9/01241, 2018/31/D/ST9/02852, and by Polish Ministry for Education and Science under contract MEiN/2021/2/DIR.



\bibliographystyle{aasjournal}
\bibliography{iplbib}

\end{document}